# A short insight about Thought experiment in Modern Physics

*Dragoljub A. Cucić*
*Regional Centre for Talents "Mihajlo Pupin"[1], Pančevo, Serbia, cule@panet.rs*

**Abstract**

Thought experiment is an apparition in a modern Science, which is in last 100 Years took sweep. Work is based at the aspect of Physics, because and the writer is from Physics. Manners to study thought experiments are multiple, and this work is only one aspect in that complex of feasibility. About real character of thought experiment will be made conclusion when analyze, like this attempt, in other Sciences.

**Introduction**

During the XX century thought experiment, as a feature of contemporary science, was extremely developed. It has been around since the Ancient Greece. Even then it was used as a scientific method (both in natural and social sciences). This work is based on physics, regarding that its author, who is physicist himself finds the best way in it. Thought experiment can be found in many sciences, and its real character is going to be discovered only when similar analyses (for example like this one) were made within each science.
There are three classes of experiment: numerical, thought and laboratory. They, all together, have some identical and some different chracteristics. Here we will talk about thought experiment.
The beginning of 20-th century can be induced as a period of thought experiment's "real" appearing in physics. Thought experiments existed before 20-th century, as it's going to be seen through the examples, but their actual expansion started when the theories of relativity and of quantum mechanics were founded. Firstly A. Einstein and then a plied of first-class physicists begun to attach much importance to the thought experiment. This period[2] is a period of huge changes in physics; a period when the basic physics' standards were revitalized; a period when scientists were looking for new recipes and when it was obvious that classical physics had been holding back an open possibility for someone to say something new in a different way.
By the end of 19-th century it was expected that classical physics and Maxwell's electrodynamics systems was about to be closed[3] and after than, during the first thirty years of this century the "world's picture" was completely changed. The term "quantum" has been brought into physics by German physicist M. Planck in 1900. A couple of years later first A. Einstein works, connected to the Special Theory of Relativity (STR) and photo effect, appeared. The base for expanding of quantum mechanics and theory of relativity has been made the limitations in making statements and new theories were a big problem for all physicists. They were not able to present theories of relativity an quantum mechanics because they didn't have appropriate "terms" to describe them with. At that time the research of these theories introduced new levels of abstractions which led towards thinking collisions completely "now set of problems" (so called "untouchable" problems) were studied. Even contemporary physics was forced to find new methods in order to solve and explain them. Stating of classical physics wasn't consistent while dealing with problem in quantum mechanics or in theory of relativity. There was a need for a "new physics" which would be somehow closer to it's manners as well as to it's users. This endless attempts to find the way of overcoming the situation which was gained by "terminological and conceptions confusion", in the anticipated scientific break, revealed thought experiment as a form of abstract thinking which enables special approach to solving, presenting and connecting of certain problems.

---

[1] Michael Idvorsky Pupin
[2] Beginning of 20-th century.
[3] Lord Kelvin, one of the leading physicists oh 19-th century, thought, that only "two litle clouds to appear on the horizon" for making general theory of physics. First one is a negativ result of Michelson-Morly's experiment for proving the existenc of ether, and the second is a failure of Rayleigh-Jeans's low to deny "ultraviolet catastrophy". Dealing with these problems brought to finding of theory of relativity and quantum theory.

A man reacts on the world around him with the help of his senses. Their number is restricted (sight, hearing, taste, smell and touch), as well as each sense's function is restricted (for example limit for sight is 400nm-700nm or for hearing is 20Hz-20000Hz). Indirect experiment is when a laboratory is carrled out in the interval if senses' limitations. Physicists of 20-th century were mainly interested in this directly perceptive world. In favor to that thought experiment was founded as one of the possibilities to find out discover something. It was founded in the same time of founding of theory of relativity and quantum mechanics which deal with physic fields that are indirectly accessible to the senses. Quantum mechanics is a part of physics which begins at the crossing from "macro" (accessible to the senses) into "micro" (inaccessible to the senses) world; the world where current laws contradict classical way of thinking about macro world. Special theory of relativity is a part of physics that begins at the crossing from "slow" (accessible to the senses) into "fast" (inaccessible to the senses). "Micro" and "fast" is "inaccessible" and this is one of the ways, except mathematics, to build up realizations of events which had happened within those domains and to imagine how would all look like. Thought experiments are those experiments which are conducted imaginative world, but not out of thoughts. In order to satisfy intuition and knowledge relations with groups of notions and familiar ideas are made.

They are different believes of what in fact is a thought experiment. In this work the author supports the idea of an Austrian physicist and philosopher E. Mach who thinks that a thought experiment is any experiment that hasn't been realized yet. There is a pattern explained here which will help us to prove of the experiment is thought experiment or not.

## 1. About an experiment

Experiment, itself, has been frequently exploited topic in many discussions of physicists, episthemologists and all those who are interested in philosophy of science. But very little is said about thought experiment; about how real is it and what exactly is experimental in it; and if it isn't an experiment then what is different between an experiment and thought experiment.

It is understood that person who is conducting the experiment has an influence on conditions for its realization The person who conduct the experiment, experimentalist, is a subject who gives form and then realizes it. Experiment can be repeated which gives an effect of confirmation. Also, while setting the experiment a certain prediction s performed. Experiment results can tell us if the experiment was successful or not and according to that has been used in that experiment.

Experiment is a scientific method, used in researching, trough which we can get new information. They enable scientists to judge the researched feature. Experiment characteristics are:

1. **Instrumentality**: if there is an intention for experiment to be realized then there should be "tools" to help it's realization, instrumentality is all objects that are parts of realizing experiment conditions. They can be imaginative (if it's thought experiment) or ponderable (if it's laboratory experiment).
2. **Testability**: this characteristic enables making of distinction between experiment and passive observation. The experiment results can be confirmed either by its repeating or by conducting of another experiment. Whose results would macth the results of new previous experiment.
3. **Singularity of statement**: it is based on already conducted experiment. It has special meaning connected to the experiment and it doesn't have any intention to generalize the explanation.

The fourth characteristic explains the difference between laboratory and thought experiment:

- **realized**: experiment has empirical base when it is realized out of fiction – this is laboratory experiment and according to it's nature, it is vericitabile.
- **unrealized**: experiment doesn't have empirical base when it is not realized out of fiction – this is thought experiment and hasn't vericitable nature.

## 2. Laboratory experiment in physics

The meaning of laboratory in physics is about to be explained in this chapter, in order to make analogy of characteristics of laboratory and thought experiments.

Laboratory experiments are used both in natural an social sciences. They are: chemistry laboratory experiments, philosophical laboratory experiments, laboratory experiments in physics and so on ...

Thought experiments can also be found both in social and natural sciences. There are thought experiments in chemistry, thought philosophical experiments, thought experiments in physics (this project is discussing this group of thought experiments), and so on ...

Author is using terms of laboratory and thought in the meaning of laboratory experiment and thought experiment in physics.

Laboratory experiment is any experiment that can be physically ponderably realized. With the help of laboratory experiment observable results can be made. The results can be: expected and unexpected. According to them or following them scientists can decide if they are going to accept, correct or reject (any) theory whose prediction has been tested by laboratory experiment.

Laboratory experiment characteristics:

1. Ponderable instrumentality
2. Testability
3. Singularity of statement
4. Realization

Explanation of the characteristics given above:

1. Laboratory experiment is realized under artificial-ponderable conditions (when there an influence of a subject which means that there are ponderable instruments that help experiment conducting).
2. The realization of laboratory experiment, itself, points out the influence of the experimentalist which enables testing of the former, already given, statement.
3. They are made from results of laboratory experiment.
4. Laboratory experiment is ponderable realized experiment. Many scientists do not agree with this statement. The realization is considered to be an act of tactile observation which is a part of experiment's results.

### *2.1. Connection between theory and laboratory experiment*

Modern age physicists have brought the separation of theoretical and experimental physics. It was realized at time when none couldn't deal any more with theoretical and experimental problem at the same time. Middle age wise men, who cannot be called physicists (because a physicist today means come thing completely different) like: Nicolle Cusanski, Giovanni Benedetti, Galileo Galilei, ... menaged to study at the same time segments of physics (what was considered to be physics at that age) that included theoretical and experimental characteristics. Later on, if we study Decartus, Newton, Higgens, Ampere, Faraday, ... and at the end Fermi (last "physicists skillful enough to study theoretical and experimental physic at the sane time) we can notice enlargement of studied methods made by complexity of physics phenomenon. Quoting a great English physicist Lord Kelvin, Pjotr Leonidovic Kapica, one of the leading Russian physicists, we can nicely explain the indivisibility of theory and laboratory experiment. Talking about mechanism of relation between theory and practice he says:

> *I would like to remind scientists of wonderful Kelvin's comparation. He compared theory with millstone and experimental data with wheat that is poured into these millstones. It is completely obvious that millstone cannot produce anything worthy by themselves (theory functionates for itself). The quality flour depends on quality of weath, because bad wheat cannot give nutrition's flour. Due to this conditional situation, a condition is both needed for high quality of experiment and for realization of successful theory as well as for getting practical results. (Kapica, 1977).*

Complementarily of theory and experiment within the process for realization of physical sciences is undoubted because the theory by itself is meaningless unless it was verified by laboratory experiment. And laboratory experiment by itself doesn't show anything unless it was conducted within general view based on theoretical assumptions. Theory and laboratory experiment make symbiosis in process of realization which starts at general and goes to particular and *vice versa*.

Pierre Duhem says that a theory in physics has an aim to display and classify experimental laws and that the only way of testing the judgement of theory is comparing that theory consequences and experimental laws which were displayed and grouped by it. It's obvious that even P. Duham supports the standpoint of theory and laboratory experiment connection within the process of realization and that he makes a clear difference of their roles.

Experiment and theory are stimulated between each other and in that stimulation theory builds up a complementary symbiosis with laboratory experiment. It is impossible to imagine a theory, in contemporary physics, which functionates as a generally accepted one without being confirmed by laboratory experiment, as well as there isn't laboratory experiment which has been conducted only for itself without having any purpose to confirm or deny something. Stephen Weinberg says that:

> *There isn't any experimental data which contradict a theory.[4]*

that means, experimental fact (data) can be confirmed by a theory. If there isn't any theory then physicists need to make one (for example: making of Rutherford's atomic model; when Rutherford had make an laboratory experiment by which he wonted to confirm the authentic of Thomson's atomic model, but the results of experiment forced him to make another model). According to this we can conclude that there is a mutual connection of experiment and theory which enables whipping of sharp boundaries that were imposed on while stating experiment characteristics and theory components.

### 3. Thought experiment in physics

There are different options about what is a thought experiment. According to some authors, thought experiment is any experiment which cannot be principal realized in practice and others think that thought experiment lasts until it was physically realized (Mach, 1905). In this work thought experiment is considered to be any experiment which has not been conducted out of mind.

There are two reasons for non-physical realization of thought experiment:

- Experiment isn't physical realized because it's realization is still in process.
- Experiment isn't physical realized because it cannot be realized.

We make distinction in "realization" of thought experiments:

- **tehnical unrealization** – when there is technological weakness and laboratory experiment can't be realized at the moment of its setting.

---

[4] Stephen W. on awarding of Nobel Prize.

- **principal unrealization** – is when thought experiment is formed under the conditions that do not match real-physical system.[5]

In scientific literature we can find thought experiment explained as: a logical operation (P.E. Sivokonj), as a theoretical judging with a form (P.V. Kopnin), as heuristic element in process of scientific realization (B.S. Dinin), as illustratively meaningful function in a process of explaining (A.M. Maleshina), as fiction without being dependant on reality (abstract thought experiment), J.R. Brown insists on pictural characteristic of thought experiment as an essential one. (Haggqvist, 1996, p. 15.)

Both thought and laboratory experiments have the form of an experiment. They have same characteristics. For example, Soren Haggqvist says that thought experiment:

> ...most deserve the epithet "experiment", because they aspire to adjudicate theory choice. (Haggqvist, 1996, p. 15.)

When we say that thought and laboratory experiments have mutual form we think that they were derived following the same "script". The fourth characteristic presents difference between thought and laboratory experiments.

Thought experiments characteristics:

1. Imaginative instrumentality
2. Testability
3. Singularity of statement
4. Unrealization

Explanation:

1. Thought experiment is formed under artificial conditions. In order to "realize" thought experiment we need to imagine "instruments" by which we can "realize" it.
2. Testability of thought experiment reflexes on the intention for testing of its final results. There is a possibility for having thought experiment which can give us some conclusion (results).
3. We predict "particularity" of both thought and laboratory experiments. This prediction is made by experimentalist according to the theoretical anticipation of results.
4. If there isn't any possibility for making conditions for conducting an experiment, or if an experiment hasn't been realized yet, then we call that kind of experiment a thought experiment.

---

[5] There is a special sub-type of principaly unrealized thought experiment which is not going to be described in this work. These thought experiment are absurd experiment and they describe experiment thought speculations that aren't connected of all the physical system – reality.

## 3.1. About mutual and different features of laboratory and thought experiments

Both laboratory and thought experiments have experimental form which is defined with experiment characteristics: instrumentality, testability and singularity of statement.

We can differ laboratory and thought experiment if we add another characteristic to each experiment:

- physical realization – to the laboratory experiment;
- physical unrealization – to the thought experiment;

Explanation:

<u>Instrumentality</u> – in laboratory and thought experiment instrumentality is manifested in a different way because in thought experiment instruments are imagined and in laboratory experiment instruments are ponderable. Treating instruments in fiction is based on treating instruments in reality.

<u>Testability</u> – Is a mutual characteristic and it is manifested in the same way by repeating or making of new experiments which will confirm the results of previous experiment.

<u>Singularity of statement</u> – it is related only to the results of both experiments.

Laboratory experiment is ponderably realized, this implies its physical reality (it can be sensually indetified). Thought experiment is imagined and it can only be realized in thoughts when we claim that laboratory experiment can be tested we think that there is a possibility for testing by measuring. Laboratory experiment helps testing of theory; it's gives the final conclusion of the experimented theory.

Theory cannot be verified by a thought experiment. The attributes, in analogy with familiar sensual domain, are used in thought experiment. We copy elements and relations which are identical with analogical case of what is already known in physical-ponderable world. In thought experiment we use deduction and analyses. We rely on the laws of physics and theoretical assumptions. We decide on the "realization" of thought experiment according to the parameters that have been accepted.

## 3.2. Types of thought experiments

There are various opinions about the usage of thought experiment according to their cause of founding, form, way of using we give following types of thought experiments:

A. auxiliary
B. Frontier
C. Realized-qualitative
D. Unrealized-qualitative

A. There is unrealized experiment, which is used for producing additional "picture" for completing the realization of a theory in physics or a law. Examples:
1. Twin paradox;
2. Archimedes lever;
3. Ray's turning in relativistic elevator;

4. Courageous astronaut at the horizon of collaborating star with a blinking flash in his hand;

B. Frontier thought experiments are imagined in frontier physical conditions that represent frontier values of specific law in physics. Example:
   1. The Carnot's ideal heat engine;

C. Realized-qualitative thought experiments can be ponderably realized but the intention for their realization is in heuristic dilemma. They show quantitative factor of change. Examples:
   1. Schrodinger's cat;
   2. Heisenberg's group of atoms with two magnets and two observers;

D. Unrealized-qualitative thought experiments have been unrealized at the moment of their devising. They are also in heuristic dilemma which points out qualitatively different way of thinking. Examples:
   1. Dr. Erenfest's flea circus;
   2. Heisenberg's microscope;
   3. Einstein-Bohr's box with a hole and a clock;
   4. Superman and a mirror;
   5. EPR paradox;
   6. Maxwell's demon;

If we stady expressive characteristic of thought experiments we can divide them into two categories:

- presentative;
- model dilemma;

Presentative thought experiments help making valid presentation of the studied theory. They stick to that theory and represent it. Presentative experiments helps presentation of conceptual closed theory consequences (for example Archimedes lever, Twin paradox, The Carnot's ideal heat engine, ...) it doesn't mean that the theory, in which the thought experiment was formed, didn't have any dilemma about its content. This experiment doesn't include these or any dilemmas.
Model dilemma thought experiments help easier studying of general terms in physic, they enable choosing of epistemological model in studying of observed problem (for example EPR experiment, identically of gravitation and inertial mass, Schrodinger's cat, ...). This kind of thought experiment points out situation where a dilemma is formed.
With development of physic and technology many of thought experiments started to be ponderably realized and because of that, today, they belong to the group of laboratory experiments J.R. Brown and L. Infeld and many other authors think that thought experiments are principally unrealized. Basic assumption in this work is that "realization" is the primal factor that separates thought experiment from physically realized laboratory experiment. Main trait of the most realized thought experiments is that they are realized from certain approximate point which satisfies the limit accepted error, for example: twins paradox (it has not been accomplished that relativistic effects are seen on a man, but they have been seen on the elementary particles)

## 3.3. Thought experiment and theory

It is necessary for us to establish a clear relation between thought experiment and theory (there is correspondence between them). That means that we need to express ourselves as precisely as it is possible. When we are talking about the meaning of thought experiment and theory in order to establish what is mutual and what is different between them. Only in some bigger work, which enables particularity of thought experiment and complex terminological structure of theory to be included, we can see how much thought experiment is a part of theory and how close it is to the laboratory experiment.

Thought experiment can only be realized in thoughts. It doesn't have methodological power to verify or deny the theory which it generates from. It doesn't have empirical base. It is directly connected to the theory and it's assumptions, conditions and experimental flow are developed. Every theory is a result of fiction which is looking for it's confirmation in laboratory experiment. Thought experiment is one of the methods in theory that helps:

- **clearer theory presentation**: twins paradox, courageous astronaut at the horizon of event with a blinking flash in his hand, ...
- **pointing out the basic theoretical premises**: Heisenberg's microscope, Maxwell's demon, ...
- **denying of formal theory results**: Einstein-Bohr's box, EPR paradox, ...

Statement, as a thought experiment result, has singular nature because of which thought experiment isn't generalized and this is one of the basic theoretical characteristics. Due to this thought experiment can be only considered as of the "elements" which "enables" theory. Thought experiment is a part of theory, it is only experiment by its name. If there aren't any theoretical laws, that are regulating relations among object and thought experiments, then thought experiment cannot be conducted. Thought experiment completes theory, it enables realization of its consequences in a "special" way. In difference to laboratory experiment which is complementary with theory, thought experiment is a part of theory. Thought experiment helps presenting some segments of theory.

## 3.4. Importance and role of thought experiment in physics

We can find thought experiment quite often in physics. We could come across even before 20-th century. The frequency of its appearing is after all the consequence of human need to present apparitions visually. man makes fiction with the intention and they are in fact visual presentations of assumed events. He managed to reach the authentic of attitudes, but wasn't able to test himself in a different way which brought him to this way of thinking: "what would happen if it happened". Starting and limiting conditions were made by a subject himself, and they were based on his experience.

The meaning of thought experiment is often denied because of it's unverificitability. Physics is a science which confirms results by measuring. thought experiment, as a part of theory, does not fulfil a conduction of evidence in physics. It is only one of the methods for presentation of a theory.

Experimentalist "specifies" precisely all conditions, he makes analogies with real-physical system (there aren't unknown system parameters) and he establishes theoretical assumptions which conducting of the experiment and gives results that should be discussed later on.
Internal consistency of thought experiment is defined by:

- **starting and limiting conditions**; they are based on the foundation of physical-real system which means that there shouldn't be any contradiction among thought experiment and laws in physics.
- **epistemological model**; that defines theory structure within which thought experiment is formed.

Reaching measurable dimensions isn't the aim of thought experiment making. That is achieved with formulae on which the theory was built up. It's aim is to point out the significant theory characteristics. There are cases when thought experiment is only an illustration of theory (for example, twins paradox). Parameters number that defies system in thought experiment is less than parameters number that defies system in laboratory experiment.

Thought experiment is non-objective thought, based on acquired knowledge, which enables easier approach to the problem (for example, Heisenberg's microscope is used for clearer presentation of Heisenberg's Uncertainty principle). Fiction of thought experiment is precisely justified and dynamic. In some thought experiments there is only theoretical justification of event development (there are premises in theory which aren't verified empirically). thought experiment needs existence of two kinds of conditions which "were made a man" and "in which he controls experiment development". They should have the same reaction as relation between experimentalist and experimental facts in laboratory experiment.

Theory cannot be verified or denied by thought experiment, but thought experiment can help explaining of certain problem, if it hasn't been done in apologetic way. We can't measure in thought experiment, we can't assume results by it, but we can get it with deduction. It can't prove anything but it has a power of presenting. It is a synthetic product of a human mind, based on simplified reality (parameters number which defies the system is reduced). Thought experiment is an abstraction, based on theory, philosophical premises and laws in physics. It helps describing of physical system behavior. It is one of the projection of laboratory experiments. It is a thinking speculation which was founded on empirical statements.

Intrigularity of thought experiment is in the fact that most of scientific challenges and problems in physical understudying of nature have been studied through thought experiment in 20-th century. As well as through intellectual cleverness which has been trying to present "serious dilemmas" and realizations in a "trivial" way. The "popularity" of thought experiment amongst 20-th century physicists was large and it was often used for explaining and criticizing of new theories. It has become a very important part of teaching in physics.

Motivating character of thought experiment is to enable clearer view of problematic feature than it would be possible by using formulae and philosophical approaches that the theory was based on. Thought experiment explains crucial dilemmas which appear at their generalizing. Frequent heuristically content in thought experiment "draws" towards better and faster problem solving.

## 4. Examples

The examples of thought experiment in physics, which are given below, belong to the group of more familiar ones[6]. You won't find any context explanation of thought experiment because that topic is so wide that another work can be written. The author has tried here to verify or deny thought experiment with the using characteristics. They help analysis of every single example described below.

Attitudes towards thought experiment are various and this project is trying notice that there isn't an explicit definition of thought experiment but there are many of its characteristics, which are the starting point for making its classification to its usage in physics.

---
[6] Reader can find more of them in postgraduating project written by D. Cucic, who has written this work as well.

## 4.1. Equivalence of gravitational and inertial mass[7]

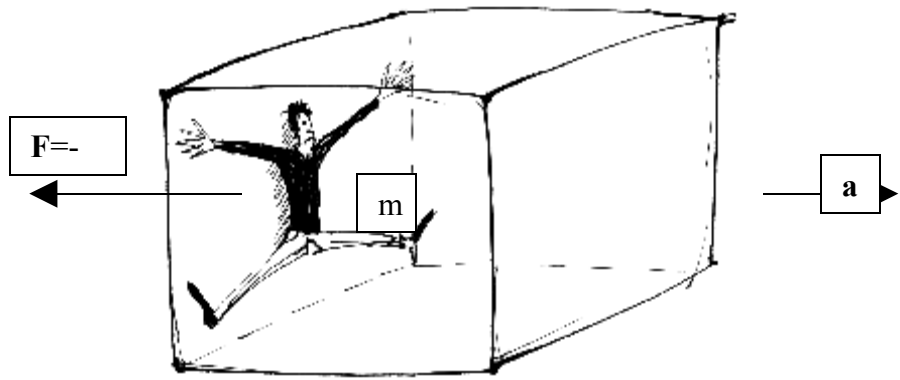

Inertial force

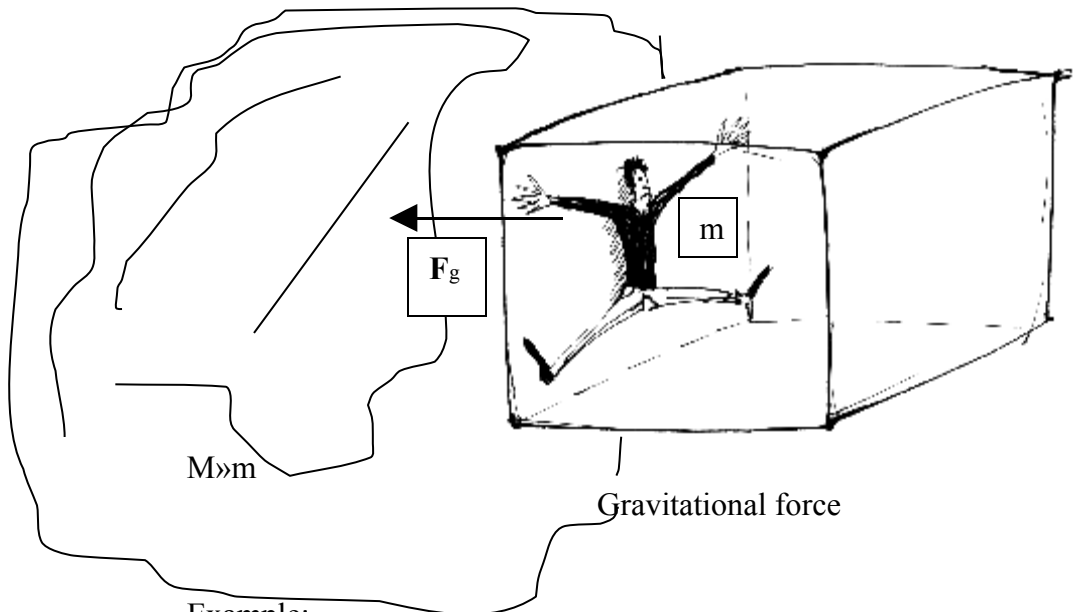

Gravitational force

Example:
Imagine a container without windows. It is in space on a sufficient distance from all masses so that there almost aren't any gravitational effects of other bodies. Imagine yourself inside the container. Imagine that you are very close to one of the container's walls.

If you know the laws in physics, and you are inside the container then there are three possibilities of your way of thinking in that situation:

1. Container is moving in a circle. The center of circumference (container's trajectory) is in imaginative direction, ortogonaly from us, towards the opposite wall of the container. We are pushing the wall with the inertial force that influences in the opposite direction from the direction of radial acceleration.
2. Container is moving in a straight line with the acceleration **a**, it is moving from us towards the opposite container wall. Using internal force (**F**=m**a**, m is our mass), we are pushing it. Force influence is the opposite direction from container acceleration.
3. There is a big mass M, on the other side of the container wall. We are standing next to it (on internal side of container) that mass has drawn us to the wall by gravitation. We influence the container wall with the force of gravitational nature.

---
[7] 1907 A. Einstein mentiones equivalency of gravitational and inertial mass for the first time.

According to Newton's law of gravitation mass has a characteristic of matter and represents a measure of mutual gravitational attraction of bodies. Inert mass presents the measure of body inertia – body tends to keep its state (if the mass is bigger than body tends to keep it's state more). There isn't any possibility to know out what kind of force, gravitational or inertial, influences on a person inside the container. According to this we can conclude that gravitational mass is equivalent to inertial mass.

Analysis:
1. Thought experiment is based on a imaging of a container without windows and a person, who knows laws in physic and is able to draw conclusions about the situation inside the container.
2. Testability can be realized because it is possible to imagine the conditions under which we can come to a conclusion about the equivalency of inert and gravitational mass.
3. The statement of "experience" of the person who is in the container is singular. It is based on assumed impression that the person inside the container has.
4. Experiment cannot be principally realized. It is impossible to bring system into a state where the complete forces influence would be zero. Today, it is technically possible to bring the system into a state in which the forces influence would be neglected enough so that we can notice the effects of experiment

According to the analysis described above, this is thought experiment.

The experiment about equivalency of gravitational inertial mass can be described as *realized qualitative* thought experiment.[8] It points out the principle of equivalency which is presented within the equivalency of gravitational and inertial mass.

This is a *model dilemma* because it shows qualitatively different understanding of mass. There is a situation which brings to dilemma what is equivalent and what is different between gravitational and inertial mass.

This thought experiment belongs to the group of *realized thought experiments*. The very day when man reached space he was able to check (very precisely how much of the abstraction matches the reality.

---

[8] Today, it is possible to realise this kind of thought experiment, but it cannot be done at the moment of its setting.

## 4.2. Equivalence between state of rest and state of uniform straight-line moving

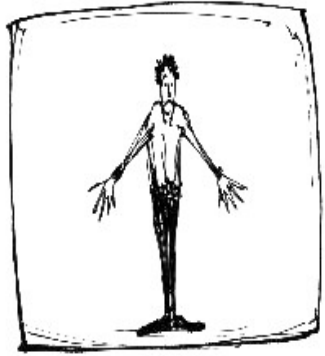

Example:
Imagine that you are in a container without windows. It is in space at a sufficient distance from any masses so that the influences of gravitational forces are negligent. You are flying inside the container and you know the laws in physics. In this situation you can think in following ways:
1. The container is resting at a sufficient distance from all masses because all gravitational forces influences are negligent.
2. The container is in state of uniform straight-line moving at forces influences are negligent.

If you imagine yourself inside the container than you realize that you cannot feel what state the container is in – state of rest or state of uniform straight-line moving. These two states are one state called unaccelerated state idest a system which has a state of inertial system. Terms of rest and uniform straight-line moving have sense only if a referential body is defined. According to this body we can observe both states.

Analyses:
1. Thought instrumentality is in fictive container and a person inside it.
2. Testability can be realized by re-imagining of conditions for the event realization.
3. Assumed judgement about experience inside the container is a special statement of imagined person inside the container.
4. Principally it is impossible for a final force influence to be equal to zero, but we can make conditions, approximately sufficient, in order to notice the effect.

In regard to all this, we can say that we were describing thought experiment.
It was possible to talk about the identity of these two states but only when the relativity of referential inertial system had been confirmed. Something – observed from one referential inertial system – is resting; but when it is observed from another system it is evenly moving.
This is *realized qualitative* thought experiment.[9] Its exceptionality is in the identity which can't be reached until some of thought terms and attitudes haven't been sufficiently explained.
This is a *model dilemma*, because it is illustrating the equivalency of two, at first sight, different states. There is a situation which brings a person inside the container into dilemma. Hi is confused about containers moving or non-moving. In regard to this dilemma we come to conclusion about the character of moving.

---
[9] The case in this example is identical to the previous one because at the moment when it has be set there wasn't any possibiliti over in to space and for precise claiming what is supposed to be aresult.

This example belongs to the group of *realized though experiments*. It was realized at the moment when we become able to sent a man into orbit.

## 4.3. Courageous astronaut at horizon of the event of collapsing star with a blinking flash in his hand

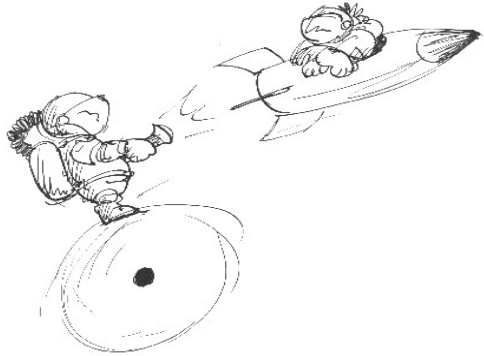

Here we are observing the influence of gravitation on light ray. Two different views on this situation are formed (following the perspective through which this event is observed).
Example:
Imagine an astronaut who is at horizon of collapsing star. He holds blinking flash and is signaling every second to another astronaut who is at the safe distance from the collapsing star. At a certain moment the collapsing star will reduce itself under critical diameter. Then, the gravitational attraction will become so strong that nothing will be able to leave its horizon of event.
As the moment of star collapsing is approaching the signals and pauses among them (they were produced by the astronaut at horizon of the event) become longer and longer for they seem that way to another astronaut who perceives them from a safe distance. They seem or are longer to him because of the influence of gravitation which collapsing star has on a light ray. The astronaut who perceives signals from a safe distance is going to see light ray at the moment of his crossing the horizon of event away from "the suicidal astronaut" and this crossing last endlessly long.
"Courageous astronaut" who dared to go to the horizon (of event) of collapsing star measures the very moment of crossing the horizon of event and intrusion into singularity within time intervals of 10000 parts of second.
Depending on referent system of observing the event, there are following situations:
If we are observing from the position of "courageous astronaut" we can see that he has already died, and that death happened at the very moment of crossing
If we are observing from the position of "astronaut at safe distance" we can conclude that "courageous astronaut" will never cross the border of the horizon (of event). (Oppenhaimer, 1967)
Analyses:
1. Thought instrumentality of this fictive event is shown in astronauts, in black hole, in blinking flash, in racket...
2. Testability is realized in imagine of conditions for event conducting and knowing of general theory of relativity.
3. Astronaut's statement, the astronaut who is at a safe distance, has a particular nature as well as a statement of a suicidal experience given by the astronaut at the horizon (of event).
4. Unrealization is both principal and technical. It is principal because there isn't a clear prove about the existence of black hole (without taking into consideration that objects, which are behaving in the same way as theory which predict black holes, are noted in space). It is technical because the existence of black holes can even be accepted but they cannot yet be reached.

According to the analyses this is thought experiment. It is *auxiliary thought experiment* because of lucid and memorable presentation of the general theory of relativity and its consequences and of black holes as product of that theory. This is *presentative thought experiment*. It points out the possibilities of general theories of relativity consequences in effective way.

*4.4. Shrodinger's cat*

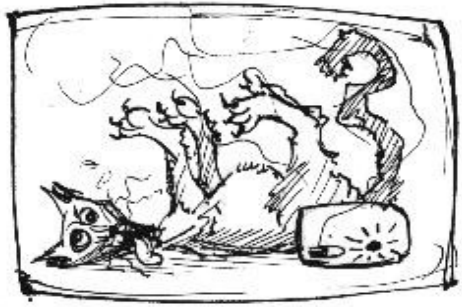

Example:
Imagine a box. Imagine: radioactive source, Geiger-Miller's counter, glass ampoule with a poison and a cat inside the box. The devise inside the box is adjusted so that detector is switched long enough in order to produce 50% chance for one of the atoms in radioactive material to fall apart which will be indicate on detector, as radioactive particles. If detector registers such event then glass ampoule with poison breaks and cat dies; but if Geiger-Miller's counter does not register radioactive particles the cat will still live.

After a certain period of time, the observer who had put the cat into box cannot know whether the cat lives or not. It is a question of statistical safety if the breaking of ampoule happened and if the poison was released due to radioactive decaying. A cat lives approximately 20 years but the experimentalist is not in position to know how long cat can live inside a box. Its fact that cat was put inside the box, which contains poisonous ampoule with radioactive trigger. Another fact is the experimentalist cannot know or define the objective reality inside the box. So we can conclude like this:
- The cat is alive
- The cat is dead
- The cat is both dead and alive

Analyses:
1. Thought experimentality is presented by imagining: of box, cat, and ampoule with poison, radioactive material...
2. Testability is realized by imaging of the events and their consequences.
3. Statement is singular because each assertion about cat draws certain new assumptions.
4. It has already been said that this experiment was realized and that it wasn't the aim to kill the cat but to try thinking in a different way. If the box isn't opened dilemma for setting this experiment is still the same so it isn't important if the experiment was realized or thought.

This experiment dosn't neither verify nor measures anything and because of that it hould be classified among thought experiment. It has in itself qualitative characteristic, which leads to chosing between two ways of thinking. Copenhagen's school of quantum mechanics supports one and the other one is supported by A. Einstein and in this case presented by E. Schrodinger. The experiment is a thought experiment. Schrodingers cat is a realized qualitative thought experiment. It means that in principle it can be realized but that isn't as much important as its characteristic is. This characteristic is reffered to qualitatively different way of reality realization, which contradicts the way, which was the cause for the experiiment.

This is an example of thought experiment, which is model dilemma. It brings suspicion into the attitude about singularity of a statement. It points out the restricted possibility of realization.

**5. Conclusion**

This work has not yet been concluded. A certain number of topics are left to be discussed about. Most of them are connected to the title of this work. The author didn't explain them because it would make his work too wide and complicated. He wanted to induce you to think about this feature physics. The big problem was non-existence of suitable terms, especially when he was trying to classify and differentiate types of thought experiments. His classification may not be the most appropriate and precise one but he imagined this work as abase on which new ideas and attitudes could be build up. First of all this work is important because it tried to clarify and systematize one unjustifiably neglected and margined method for observing physics. There is a tendency here for pointing out those "invisible threads" which connect and differ thought experiments in physics and which classify them into certain groups.

Even with more developed technology, which widens up ponderable realizations of experiments, thought experiment has its own important role as one of auxiliary methods for researching in physics. Its expressive power is doubtless. It enables easier understanding of concepts and ideas; it helps visualization of "numbers and letters". Thought experiment doesn't verify conclusions (in the sense of scientific theory). It is a non-relevant form of relevant content. Non-relevant form appears because of its inability to be proved. All the questions and dilemmas, which are used in thought experiments, are very relevant. Beside already mentioned factors of thought experiment there are others: intuitive, presentative, heuristic and illustrative which point out big importance of thought experiment as inseparable element in the process of knowledge.

Every experiment that isn't ponderably realized is in fact thought experiment. At the very moment when experiment is realized thought experiment becomes laboratory experiment. The contents of thought experiment are mostly based on experience, they are metaphysical. Thought experiment is strictly built on mathematically defined knowledge whoch gives more specified significance to their metaphysical character. Thought experiment doesn't fly through air, it has contact with earth. Thought experiment is a connecting body. It connects fictional and experienced. When experience is noticed within the form of fictional then we have thought experiment.

It is imagined as if it is going to be realized by using the experience and theory (where it was founded) and lateron results, which are developed out of previous two elements. Results are reached not direct "reading" of laboratory experiments but by "assumed" reading of fictive instruments. Thought experiment represents a result of translating theoretical terminology into the language experience where the premise of unrealization is adopted.

**Table**

| THOUGHT EXPERIMENT | AUXILIARY | FRONTIER | REALIZED QUALITATIVE | UNREALIZED QUALITATIVE |
|---|---|---|---|---|
| **PRESENTATIVE** | *1. TWIN PARADOX*<br><br>2. ARCHIMEDES LIVER<br>3. RAY'S TURNING IN RELATIVISTIC ELEVATOR<br>*4. COURAGEOUS ASTRONAUT AT THE HORIZON OF COLLABORATING STAR WITH A BLINKING FLASH IN HIS HAND* | 1. THE CARNOT'S IDEAL HEAT ENGINE | 1. HEISENBERG'S MICROSKOPE | *1. SUPERMAN AND A MIRROR* |
| **MODEL DILEMMA** | | | *1. SCHRODINGER'S CAT*<br><br><br>2. A PILE OF ATOMS WITH TWO MAGNETS AND TWO OBSERVERS | BOHM-AHARON'S EXPERIMENT<br>2. EQUIVALENCE OF GRAVITATIONAL AND INERTIAL MASS<br>3. EQUIVALENCE BETWEEN STATE OF UNIFORM STRAIGHT-LINE MOVING AND STATE OF REST<br>4. MAXWELL'S DEMON<br>5. DR. ERENFEST'S FLYS CIRCUS<br>6. HEISENBERG'S MICROSCOPE<br>7. EINSTEIN- BOHR'S BOX<br>*8. PARADOX LUI DE BROGLE BOX'S*<br>*9.EPR-PARADOKS* |

Cursive signifiance paradox
Realized experiments are marked with till colour